\pdfoutput=1

\documentclass[11pt]{article}
\usepackage{algorithm} 
\usepackage{algorithmic}
\usepackage[]{emnlp2021}
\usepackage{multirow}
\usepackage{times}
\usepackage{latexsym}
\usepackage{graphicx}
\usepackage{microtype}
\usepackage{amsmath}
\usepackage{color}
\usepackage{algorithm} 
\usepackage{algorithmic} 
\usepackage{enumitem}

\usepackage{subcaption} 
\usepackage{amssymb}
\usepackage{epsfig}
\usepackage{graphicx}
\usepackage{times}
\usepackage[T1]{fontenc}

\usepackage[utf8]{inputenc}
\usepackage[
  separate-uncertainty = true,
  multi-part-units = repeat
]{siunitx}
\usepackage{microtype}

\newcommand{\name}{\textsc{TAG}}
\title{TAG: Gradient Attack on Transformer-based Language Models}

\author{Jieren Deng$^{1}$\textsuperscript{\textsection},Yijue Wang$^1$\textsuperscript{\textsection}, Ji Li$^2$, Chenghong Wang$^{3}$, Chao Shang$^4$, Hang Liu$^5$\\ \textbf{Sanguthevar Rajasekaran$^1$, Caiwen Ding$^1$} \\
        $^1$University of Connecticut, $^2$Microsoft, $^3$Duke University, $^4$JD AI Research, 
        \\
        $^5$Stevens Institute of Technology\\
        \small\texttt{\{jieren.deng,yijue.wang,sanguthevar.rajasekaran,caiwen.ding\}@uconn.edu}\\
        \small\texttt{\{changzhouliji,chaoshangcs\}@gmail.com, chenghong.wang552@duke.edu, hliu77@stevens.edu}\\ }

\begin{document}
\maketitle
\begin{abstract}
Although distributed learning 
has increasingly gained attention in terms of effectively utilizing local devices for data privacy enhancement,  recent studies show that publicly shared gradients in the training process can reveal the private training data (gradient leakage) to a third party. 
However, so far there hasn't been any systematic study of the gradient leakage mechanism of the Transformer based language models.
In this paper, as the \textit{first} attempt, we formulate the gradient attack problem on the Transformer-based language models and propose a gradient attack algorithm, \name{}, to recover the local training data. 
Experimental results on Transformer, TinyBERT$_{4}$, TinyBERT$_{6}$, BERT$_{BASE}$, and BERT$_{LARGE}$ using GLUE benchmark show that compared with DLG~\cite{zhu2019deep}, \name{} works well on more weight distributions in recovering private training data and achieves 1.5$\times$ Recover Rate and 2.5$\times$ ROUGE-2 over prior methods without the need of ground truth label. 
\name{} can obtain up to 88.9$\%$ tokens and up to 0.93 cosine similarity in token embeddings from private training data by attacking gradients on CoLA dataset. In addition, \name{} is stronger than previous approaches on larger models, smaller dictionary size, and smaller input length. 

\end{abstract}
\begingroup\renewcommand\thefootnote{\textsection}
\footnotetext{Equal contribution}
\begin{figure*}[t]
\centering
	\includegraphics[width=0.85\textwidth]{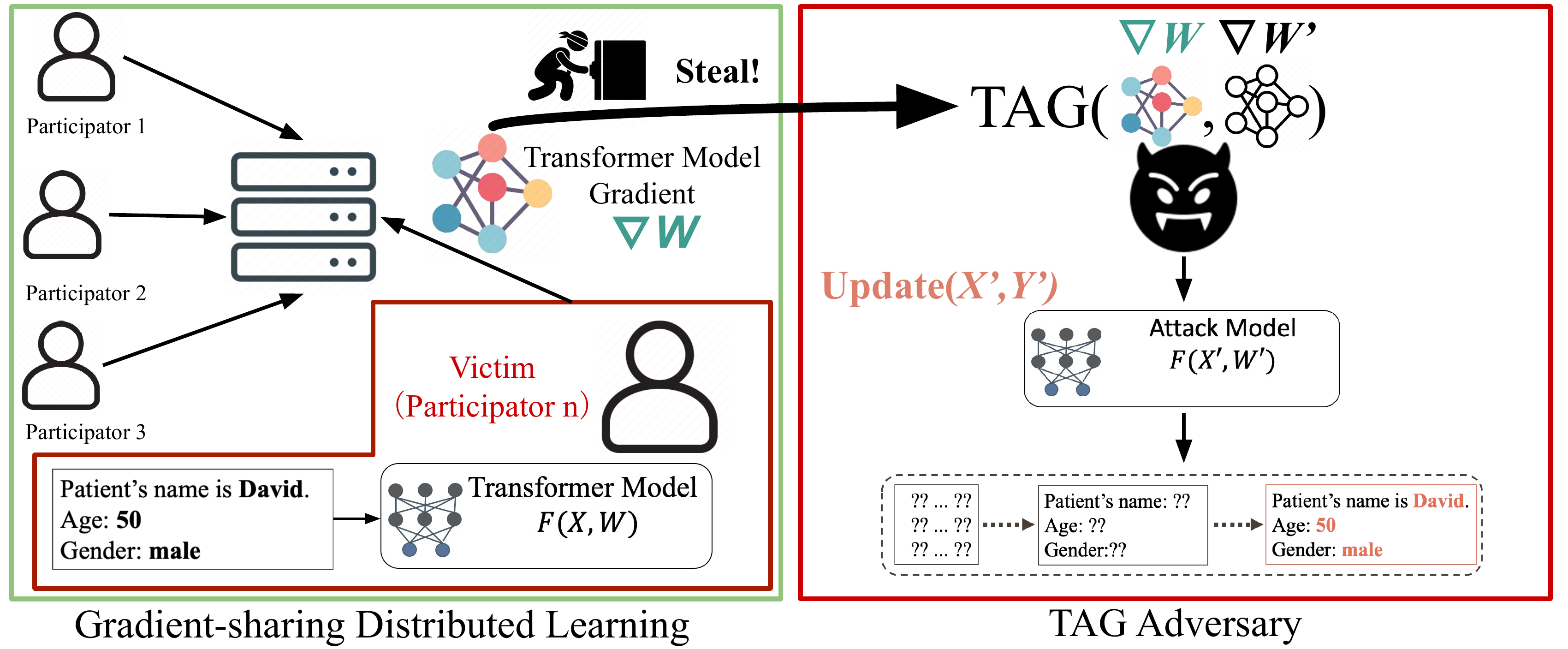}
    \caption{ Gradient transformer attack process.}
    \label{fig:intro}
\end{figure*}
\section{Introduction}
Distributed training has gained popularity in reducing training time on large-scale deep learning models and datasets~\cite{NIPS2012_6aca9700,10.5555/2685048.2685095,NIPS2014_1ff1de77,NEURIPS2019_ec1c5914,Gurevin2021Enabling,10.1093/bib/bbaa090,Lin2020ESMFLEA}.
In such systems, multiple devices or participators collaborate in training one task and 
 synchronize via exchanging gradients, allowing participants at different location for model training with their own data. 
It is widely believed that sharing gradients between participants will not leak the private
training data. 
On the other hand, 
large scale contextual representation models, such as 
ELMo~\cite{peters2018deep}, BERT~\cite{devlin2019bert}, XLNet \cite{yang2019xlnet}, T5 \cite{2019t5}, and GPT-3~\cite{GPT-3} have significantly promoted natural language processing (NLP) in the last decade. 
Thus, the ability of parallel training has helped propel using distributed learning on a large scale NLP, for efficient training. 

Recent studies show that private training data can be recovered through the deep learning model by gradients~\cite{zhu2019deep,chen2020improved,NEURIPS2020_c4ede56b, ijcai2021-432}.
For instance, recent work "Deep
Leakage from Gradient” (DLG)
\cite{zhu2019deep} showed the shared gradients could leak private training data in image classification.
Another recent work IG~\cite{NEURIPS2020_c4ede56b} shows that it is possible to recover images by gradients in a trained network. 

Despite the success, there are several 
limitations on current works: (i) Lack of generalizability on different weight
distribution. They only succeeded in the early training phase at certain weight distribution. (ii) Lack of formal problem formulation and gradient attack evaluation in the field of NLP. Existing work only show the reconstruction difference at the sentence level without quantitative analysis. (iii) There is little investigation on the impact of different attention heads, model architectures on Transformer gradient attack.

In this paper, we propose a novel algorithm,
to recover 
private training data of Transformer-based language model from the shared gradients.
As shown in Figure~\ref{fig:intro}, our TAG adversary obtains the transformer model gradients $\nabla$\textit{W} from a distributed learning participator and updates initialized dummy data (\textit{X}$^\prime$, \textit{Y}$^\prime$) by comparing the difference between the participator's gradients $\nabla$\textit{W} and adversary's gradients $\nabla$\textit{W}$^\prime$. Eventually, the adversary recovers the dummy data (\textit{X}$^\prime$, \textit{Y}$^\prime$) and acquires the information from the participator's private training data \textit{X} such as name, age, and gender. Our contributions are summarized as follows:


\begin{itemize}[leftmargin=*]
\item As the \textit{first} attempt in the field of NLP, we propose a general gradient attack algorithm, \name{}, to recover the private training data on Transformer-based language models. Compared to the existing methods, \name{} works on more realistic weight distributions, including both pre-trained models and normally initialized models.



\item We develop a quantitative evaluation method on the NLP gradient attack problem while the existing work shows the recovered texts. We use a set of metrics (Recover Rate, ROUGE-1(\%), ROUGE-2(\%), ROUGE-L(\%), and runtime) to evaluate the effectiveness of the proposed attack algorithm. With these metrics, TAG achieves 1.5$\times$ Recover Rate and 2.5$\times$ ROUGE-2 over prior methods. TAG can also obtain up to 88.9\% tokens and up to 0.93 consine similarity in token embeddings from private training data.




\item We conduct a comprehensive analysis of different weight distribution, dataset, vocabulary dictionary size, and model size on Transformer,  TinyBERT$_{4}$, TinyBERT$_{6}$, BERT$_{BASE}$, and BERT$_{LARGE}$, and we observe that \name{} has a stronger adversary on large models than on small ones. In addition, models with a smaller dictionary size and smaller input sequence length are riskier in leaking the private training data. 


\end{itemize}

\section{Related Work}\label{sec:related works}
\subsection{Privacy leakage problem}
Privacy leakage is studied in the training phase and prediction phase. Privacy attack from gradient and model inversion (MI) attack \cite{fredrikson2015model} aim at the training phase by constructing the features of the private training data by using the correlation between the private training data and the model output. 
The authors in \cite{fredrikson2015model} showed that it is possible to infer individual genomic data via access to a linear model for personalized medicine. Recent works extend MI attack to recover features of private training data of Deep Neural Networks (DNNs). Privacy attack from gradients is different from previous MI attack. It recovers the private training data exploiting their gradients in a machine learning model. The process of privacy leakage from gradients is shown at Figure~\ref{fig:intro}.

\subsection{Distributed learning}

Distributed learning is a popular framework for large-scale model training~\cite{das2016distributed, NIPS2012_6aca9700,10.5555/2685048.2685095,NIPS2014_1ff1de77,NEURIPS2019_ec1c5914} that leverage the computation power of many devices by aggregating the local models trained on the devices.
Instead of training a model with all the data at a server, each device trains a local model with a different chunk of the dataset and shares the final gradients. 
A popular distributed learning algorithm is \textit{Synchronous Stochastic Gradient Descent} (sync-SGD)~\cite{10.5555/2685048.2685095,NIPS2014_1ff1de77} which contains a single server and $n$ local devices. Each device trains a local model and shares the gradient with the server. The server then aggregates the gradients of the different devices and starts another round by sharing the aggregated result with the devices.

\subsection{Prior arts on gradients-based attack}
Although a distributed learning system protects privacy by not sharing private training data, research works have shown that it is possible to infer the information of private training data from the shared gradients in both language tasks and computer vision tasks. ~\cite{melis2019exploiting} enables the identification of words used in the training tokens by analyzing the gradients of the embedding layer. \cite{goodfellow2014generative} proposes an attack algorithm to synthesize images mimicking the real training images by Generative Adversary Network (GAN) models.  
Besides the works that recover certain properties of the private training data, DLG~\cite{zhu2019deep} is a more recent work that shows that it is possible to recover private training data with pixel-wise accuracy for images and token-wise matching for texts by gradient matching. 
DLG~\cite{zhu2019deep} achieves the recovery of images from different datasets on LeNet-5. However, DLG~\cite{zhu2019deep} has limitations on evaluating the performance thoroughly on different weight distribution settings, various networks, and different training stages (pre-trained versus initialized). 
To the best of our knowledge, there is no existing work that comprehensively investigates gradient-based attacks for transformer-based language models with benchmark dataset and standard metric. 



\section{Approach}
In this section, we first formulate the gradient attack in NLP, and the proposed algorithm is introduced afterward. 

\subsection{Transformer-based NLP models}

Transformer~\cite{vaswani2017attention} is the fundamental architecture for many popular pre-trained language models, e.g., BERT~\cite{devlin2019bert}. 
Scaled dot-product self-attention is the underlying key mechanism inside Transformer, which is calculated as 
\begin{equation}\label{eqn:singlehead}
\small sdpsAttention(q, k, v) = v\cdot softmax(\frac{{q\cdot k}^T}{\sqrt{d_k}})
\end{equation}
where $q$, $k$, and $v$ represent the query, key, and value, respectively, and $1/\sqrt{d_k}$ is a scaling factor. 
Multi-head attention is applied to first calculate attention using Eq. \ref{eqn:singlehead} in the subspace of embeddings and then concatenate to form the final output. 

A typical flow is to first pre-train the Transformer with objectives like masked language modeling on huge amounts of unlabeled data to get a pre-trained model like BERT~\cite{devlin2019bert} and RoBERTa~\cite{liu2019roberta}, and then finetune the pre-trained model on specific downstream tasks using the labeled data. 

In spite of the high accuracy achieved by the Transformer based models, the large size and high latency of such models make them less appealing to resource constrained edge devices.
Accordingly, various knowledge distillation and model compression techniques have been proposed to effectively cut down the model size and inference latency with minimum impact on the accuracy.

Without any loss of generality, we consider the Transformer~\cite{vaswani2017attention}, BERT~\cite{devlin2019bert}, and TinyBERT~\cite{jiao2020tinybert} as the representatives of encoder-decoder Transformers, decoder only pre-trained large Transformers, and compressed pre-trained Transformers. Our approach can be extended to other similar models, such as RoBERTa~\cite{liu2019roberta}, UniLMv2~\cite{bao2020unilmv2}, and DistilBERT~\cite{sanh2019distilbert}.

\subsection{NLP gradient attack formulation}
We assume that an adversary cannot access the private training data $({\mathbf{X}}, {\mathbf{Y}})$ in local training directly, but the adversary can gain the gradients that the local devices shared, and the current global model $\mathcal{F}(\mathbf{X},\mathbf{W})$ in any timestamps during training, where $\mathbf{X}$ is input tokens and $\mathbf{Y}$ is the output labels and $\mathbf{W}$ is the model weights.

The objective of the attack is to recover the valuable and private training data using the shared gradients. For computer vision models, the objective is to retrieve the original pixels in the training images. As mentioned in Section \ref{sec:related works}, most prior works fall into this category. 
In this work, we focus on modern Transformer-based models for NLP applications, and our goal is to recover the original tokens in the training set. 

Attacking NLP applications is more challenging than computer vision applications, and the reasons are threefold. 
First, the range of possible values at each pixel is usually smaller than the range of possible tokens at each position, and it is generally more difficult to find the exact item from a larger candidate space. Second, the valuable information carried in an image can be retrieved from a region of pixels, whereas for NLP data, the sensitive information could be carried by several specific tokens, e.g., name, time, and location, and it is required to achieve an exact match on the tokens at certain positions to get the key information from the original text. Third, humans can tolerate more errors at pixel values in an image, whereas a small error in the retrieved token id leads to irrelevant token strings in most cases. 

Without any loss of generality, the attack can happen at any training stage of the shared global model, and we consider the two most common weight initialization methods, including random initialization for non-pre-trained models and specific learned values for pre-trained models. More formally, the problem is formulated as:
\begin{equation}
  \begin{aligned}
     \text{Constructing  } &(\mathbf{X}^{*}, \mathbf{Y}^{*})\\ \textrm{s.t.} \frac{\partial \mathcal{L}(\mathbf{W},\mathbf{X}^{*}; \mathbf{Y}^{*})}{\partial \mathbf{W}} &= \frac{\partial \mathcal{L}(\mathbf{W},\mathbf{X}; \mathbf{Y})}{\partial \mathbf{W}}
  \end{aligned}
\end{equation}
where $(\mathbf{X}^{*}, \mathbf{Y}^{*})$ are the recovered data, i.e., tokens and labels for language tasks.   

\subsection{Proposed algorithm}
\subsubsection{Recovered token initialization}
To recover the language data, we first randomly initialize a dummy data $(\mathbf{X}^\prime,\mathbf{Y}^\prime)$, where $\mathbf{X}^\prime$ is called dummy input and  $\mathbf{Y}^\prime$ is called dummy label. Then we get the corresponding dummy gradient as:
\begin{equation}
    \nabla{{\mathbf{W}}^\prime}  = \frac{\partial \ell(\mathcal{F}(\mathbf{W},\mathbf {X}^\prime); {\mathbf{Y}}^\prime)}{\partial \mathbf{W}}
\end{equation}

The next step is to optimize dummy gradient, $\nabla{{\mathbf{W}}^\prime}$, to ground truth gradient $\nabla{{\mathbf{W}}}$, as closer as possible. In this case, we need to define a differentiable distance function $\mathcal{D}({\mathbf{W}},{\mathbf{W}^\prime})$, so that we can obtain the best ${\mathbf{X}^\prime}$ and ${\mathbf{Y}^\prime}$ as ($\mathbf{X}^*$,$\mathbf{Y}^*$) follows:

\begin{equation}
    ({\mathbf{X}^*},{\mathbf{Y}^*}) = \underset{ ({\bf{X}}^{\prime},{\bf{Y}}^{\prime}) }{\text{arg min }} {\mathcal{D}}(\nabla{{\mathbf{W}}^\prime},\nabla{\mathbf{W}})
\end{equation}

\subsubsection{Distance function for gradient matching}
Our experimental observation shows that with different weight initialization, for the same private training data, ${\mathbf{X}}$, the NLP model may have distinctly different gradients, $\nabla{{\mathbf{W}}}$. For example, with initialized weights from a normal distribution, the gradient of the model may be larger in magnitude than with initialized weights from a uniform distribution (two distributions have the similar intervals). Besides, the $\nabla{{\mathbf{W}}}$ gathers near-zero values more heavily with weights from normal distribution than with weights from uniform distribution. We have a similar observation that the $\nabla{{\mathbf{W}}}$ gathers considerable near zero values with weights from a pre-trained model. We consider a matrix with substantial near-zero values as a sparse matrix.

If we use the Euclidean distance (which is used in DLG~\cite{zhu2019deep}) to measure the difference between $\nabla{{\mathbf{W}}^\prime}$ and $\nabla{{\mathbf{W}}}$, the recovery of the ground truth data
is driven by large gradients at the early stages. As a result, this
might cause a problem when using Euclidean distance under a
normal weight initialization since most of the gradients gather
around zero while a small proportion of gradients have large
values.

To overcome this problem, instead of using the Euclidean distance for $\nabla{{\mathbf{W}}^\prime}$ and $\nabla{{\mathbf{W}}}$ as the distance function, we consider a combined distance of L2 norm (Euclidean distance) and L1 norm (Manhattan distance), and a coefficient parameter, $\alpha$, 
 as our distance function to measure the difference between $\nabla{{\mathbf{W}}^\prime}$ and $\nabla{{\mathbf{W}}}$:
\begin{equation}
\small \begin{split}
     &\mathcal{D}(\nabla{{\mathbf{W}}}^{\prime},\nabla{{\mathbf{W}}}) \\ 
    &=||\nabla{\mathbf{W}}^{\prime}-\nabla{\mathbf{W}}||_2  
    + \alpha(\nabla{\mathbf{W}})||\nabla{\mathbf{W}}^{\prime}-\nabla{\mathbf{W}}||
\end{split}
\label{eq:lossw}
\end{equation}
where $\alpha({\nabla\mathbf{W}})$ is a factor specified for each layer's ${\nabla\mathbf{W}}$  and its value decreases along with the order of the layer. By doing this, we put larger weights on the gradient differences on the front layers as they are closer to the input private training data. The value of $\alpha({\nabla\mathbf{W}})$  is crucial and needs to be suitable for different weight settings.  

\subsection{The framework of the algorithm}
Our complete proposed algorithm is shown in Algorithm~\ref{alg:alg1}, and the highlights of our algorithm are as follows. 
We initialize dummy input \textit{(dummy token embeddings)} and dummy label, $(\mathbf{X}^\prime, \mathbf{Y}^\prime)$, as dummy data in line 2. 
During the iterations, started from line 3, we first obtain the dummy gradient, $\nabla\mathbf{W}^\prime$, of the current dummy input. Then we use the distance function in Eq. \ref{eq:lossw} to measure the difference, $\mathcal{D}(\nabla\mathbf{W}, \nabla\mathbf{W}^\prime_i)$, between dummy gradient $\nabla\mathbf{W}^\prime$ and ground truth gradient $\nabla\mathbf{W}$. At the end of each iteration, we update the $(\mathbf{X}^\prime, \mathbf{Y}^\prime)$ by the calculated difference, $\mathcal{D}(\nabla\mathbf{W}, \nabla\mathbf{W}^\prime_i)$ in line 7 and line 8. When a pre-set maximum number of iterations is reached, or in 200 iterations, or the number of recovered tokens in ground truth does not change, the algorithm will eventually output the optimized $(\mathbf{X}^\prime, \mathbf{Y}^\prime)$ as $(\mathbf{X}^*, \mathbf{Y}^*)$ after the iterative recovery process. 
\begin{algorithm}[h]\footnotesize

\begin{algorithmic}[1]
\STATE Input: $\nabla\mathbf{W}$: ground truth gradient; $\mathcal{F}(\mathbf{X},\mathbf{W}^{\prime})$: NLP model; $\eta$: learning rate; $\mathbf{W}^{\prime}$: parameter weights
\STATE Initial: $\mathbf{X}^\prime\sim \mathcal{N}(0,\,1)$, $\mathbf{Y}^\prime\sim \mathcal{N}(0,\,1)$
		\FOR{the i-th iteration}
			\STATE$\nabla\mathbf{W}^\prime_{i} \leftarrow \partial{\ell (\mathcal{F}(\mathbf{X}^{\prime},\mathbf{W}^{\prime})}/\partial{\mathbf{W}^\prime})$  //get dummy gradient by TAG
			\STATE $\mathcal{D}(\nabla\mathbf{W}, \nabla\mathbf{W}^\prime_i)  \leftarrow  \|\nabla\mathbf{W}^\prime_i-\nabla{\mathbf{W}}\|_2   + \alpha(\nabla{\mathbf{W}})\|\nabla\mathbf{W}^\prime_i -\nabla{\mathbf{W}}\|$
            
			\STATE \textbf{update} $(\mathbf{X}^\prime,\mathbf{Y}^\prime)$:
			\STATE
			  $\mathbf{X}^\prime \leftarrow \mathbf{X}^\prime-\eta \frac{\partial \mathcal{D}(\nabla\mathbf{W}, \nabla\mathbf{W}^\prime_i)} {\partial \nabla \mathbf{X}^\prime}$,
			\STATE
			  $\mathbf{Y}^\prime \leftarrow \mathbf{Y}^\prime-\eta \frac{\partial \mathcal{D}(\nabla\mathbf{W}, \nabla\mathbf{W}^\prime_i)} {\partial \nabla \mathbf{Y}^\prime}$
		\ENDFOR
		
\STATE Output: Recovered Data
$\mathbf{X}^*$,$\mathbf{Y}^*$
\end{algorithmic}
\caption{TAG}\label{alg:alg1}
\end{algorithm}


\section{Experimental setup}
All of our experiments are conducted on servers with Intel(R) Xeon(R) Gold 5218 (64 virtual CPUs with 504 GB memory) and 8 NVIDIA Quadro RTX 6000 GPUs (24GB memory) by PyTorch 1.5.1, Python 3.6, and CUDA 10.2.

\subsection{Datasets}\label{sec:dataset}
We evaluate our algorithm on the following tasks from the General Language Understanding Evaluation (GLUE)~\cite{wang2019glue} benchmark. 

\noindent\textbf{CoLA.} The Corpus of Linguistic Acceptability~\cite{cola2019} consists of English acceptability judgments drawn from book and journal articles on linguistic theory. Each example is a sequence of words annotated with whether it is a grammatical English sentence. 

\noindent\textbf{SST-2.} The Stanford Sentiment Treebank~\cite{socher-etal-2013-recursive-SST-2} consists of sentences from movie reviews and human annotations of their sentiment. The task is to predict the sentiment of a given sentence. We use the two-way (positive/negative, 1/0)  class split and use only sentence-level labels.

\begin{table}[]
    \scalebox{0.9}{
    \begin{tabular}{l|llll}
\hline
Models & Layers & \begin{tabular}[c]{@{}l@{}}Hidden \\ Units\end{tabular} & \begin{tabular}[c]{@{}l@{}}Attention \\ Heads\end{tabular} & \begin{tabular}[c]{@{}l@{}}Filter \\ Size\end{tabular} \\ \hline
Transformer & 2 & 100 & 4 & 200  \\
TinyBERT$_{4}$ & 4 & 312 & 6 & 1,200  \\
TinyBERT$_{6}$ & 6 & 768 & 12 & 3,072  \\
BERT$_{BASE}$ & 12 & 768 & 12 & 3,072  \\
BERT$_{LARGE}$ & 24 & 1,024 & 16 & 4,096  \\ \hline
\end{tabular}}
    \caption{Model structures of Transformer, TinyBERT$_{4}$, TinyBERT$_{6}$, BERT$_{BASE}$, BERT$_{LARGE}$.}
    \label{tab:model size}
\end{table}

\noindent\textbf{RTE.} The Recognizing Textual Entailment (RTE)~\cite{RTE-dagan2005pascal} datasets come from a series of annual textual entailment challenges. 
This dataset is constructed based on news and Wikipedia text with a combination of  RTE1-3, and RTE5. 

We select these three datasets because they contain sentences of different lengths. Typically, sentences are about 5 to 15 words for CoLA, 10 to 30 words for SST-2, and 50 to 100 words for RTE.
In fact, our algorithm is data agnostic, which can work on any text inputs from any benchmark, or even any sentence from any source.

\subsection{Model settings}
We conduct experiments using three popular transformer-based networks, including the basic transformer model~\cite{vaswani2017attention}, TinyBERT ~\cite{jiao2020tinybert} and BERT~\cite{devlin2019bert}. The basic transformer contains two transformer encoders and one transformer decoder. The number of heads in the self-attention layers is four, and the dimension of the feed-forward network model is 200. The activation function is Gaussian Error Linear Units (GELU)~\cite{hendrycks2016gaussian}. We also applied our algorithm to two different sizes TinyBERT and two different sizes BERT.  The TinyBERT$_{4}$ is with four layers, 312 hidden units,  feed-forward filter size of 1200 and 6 attention heads. The TinyBERT$_{6}$ is with 6 layers, 768 hidden units,  feed-forward filter size of 3072 and 12 attention heads. In addition, we use the configurations from \cite{devlin2019bert} for BERT. The BERT$_{BASE}$ has 12 layers, 768 hidden units,  3072 feed-forward filter size, and 12 attention heads. The BERT$_{LARGE}$ has 24 layers, 1024 hidden units,  4096 feed-forward filter size and, 16 attention heads. 
Table~\ref{tab:model size} summarizes the model structures explored in this work.

\begin{figure*}[t]
\centering\scalebox{0.9}{
         \includegraphics[width=\textwidth]{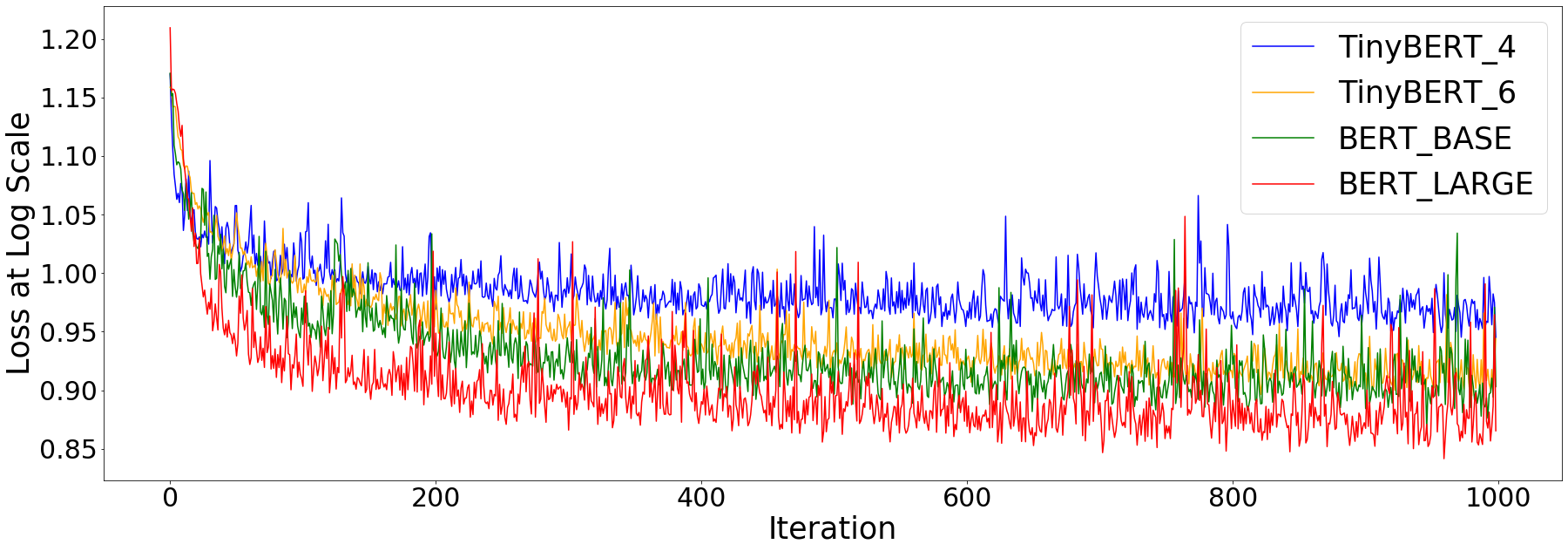}}
         \caption{The average Loss vs Iteration curve of models TinyBERT$_{4}$, TinyBERT$_{6}$, BERT$_{BASE}$, BERT$_{LARGE}$ on data CoLA, SST-2 and RTE. The loss decreases at the first 200 iterations and becomes stable after 200 iterations.}
         \label{fig:loss}
\end{figure*}

\subsection{Experiment parameters settings}
For each task and dataset of interest, we use the same set of hyperparameters: BertAdam optimizer \cite{devlin2019bert} with learning rate 0.05. For every single sentence recovering, we set the max iteration as 1,000 for our algorithm.

\subsection{Experiment evaluation}\label{sec:metric}
\textbf{Recover Rate}. This metric is defined as the maximum percentage of tokens in ground truth recovered by TAG. 

\noindent\textbf{ROUGE}. Recall-Oriented Understudy for Gisting Evaluation~\cite{lin-2004-rouge}, is a set of metrics used for evaluating automatic summarization and machine translation in natural language processing.
We use ROUGE-1, ROUGE-2, and ROUGE-L to evaluate the similarity between the sentence generated from gradient attacks and the original sentences. More specifically speaking, ROUGE-1 and ROUGE-2 refer to the overlap of unigrams and bigrams between the recovered text and reference, respectively, and ROUGE-L  measures the longest matching subsequence of tokens. 

\begin{figure*}
     \centering
      \includegraphics[width=0.9\textwidth]{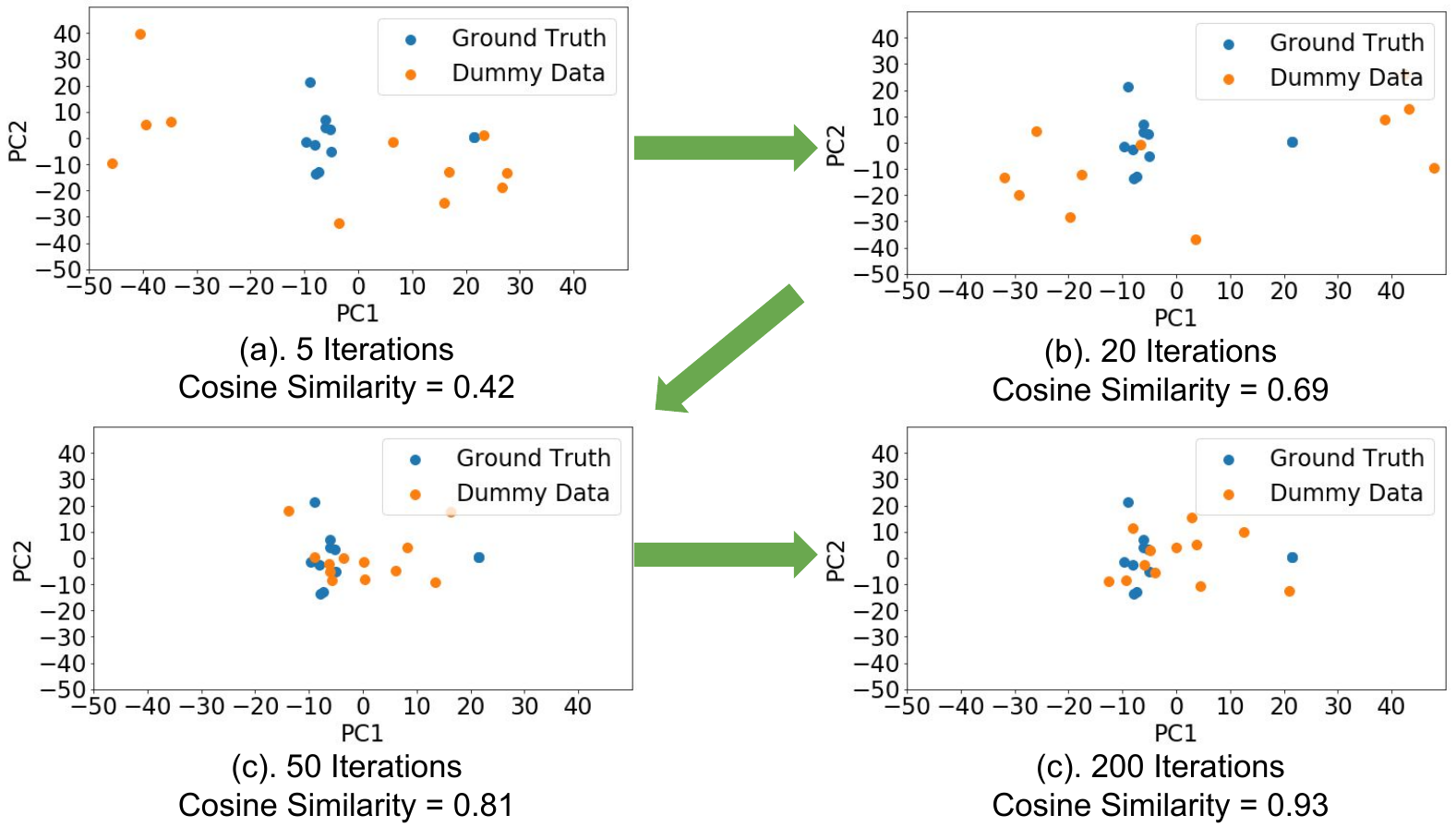}
        \caption{PCA 2D plot for dimension reduced token embeddings of TinyBERT$_{4}$ on CoLA.
       The cosine similarity of dimension reduced token embeddings between dummy data and ground truth increase with training iterations.}
         \label{fig:pca}
\end{figure*}
\noindent\textbf{Runtime}. This metric is the average of elapsed system time to complete the attack. 
\begin{figure}
     \centering
      \includegraphics[width=0.5\textwidth ]{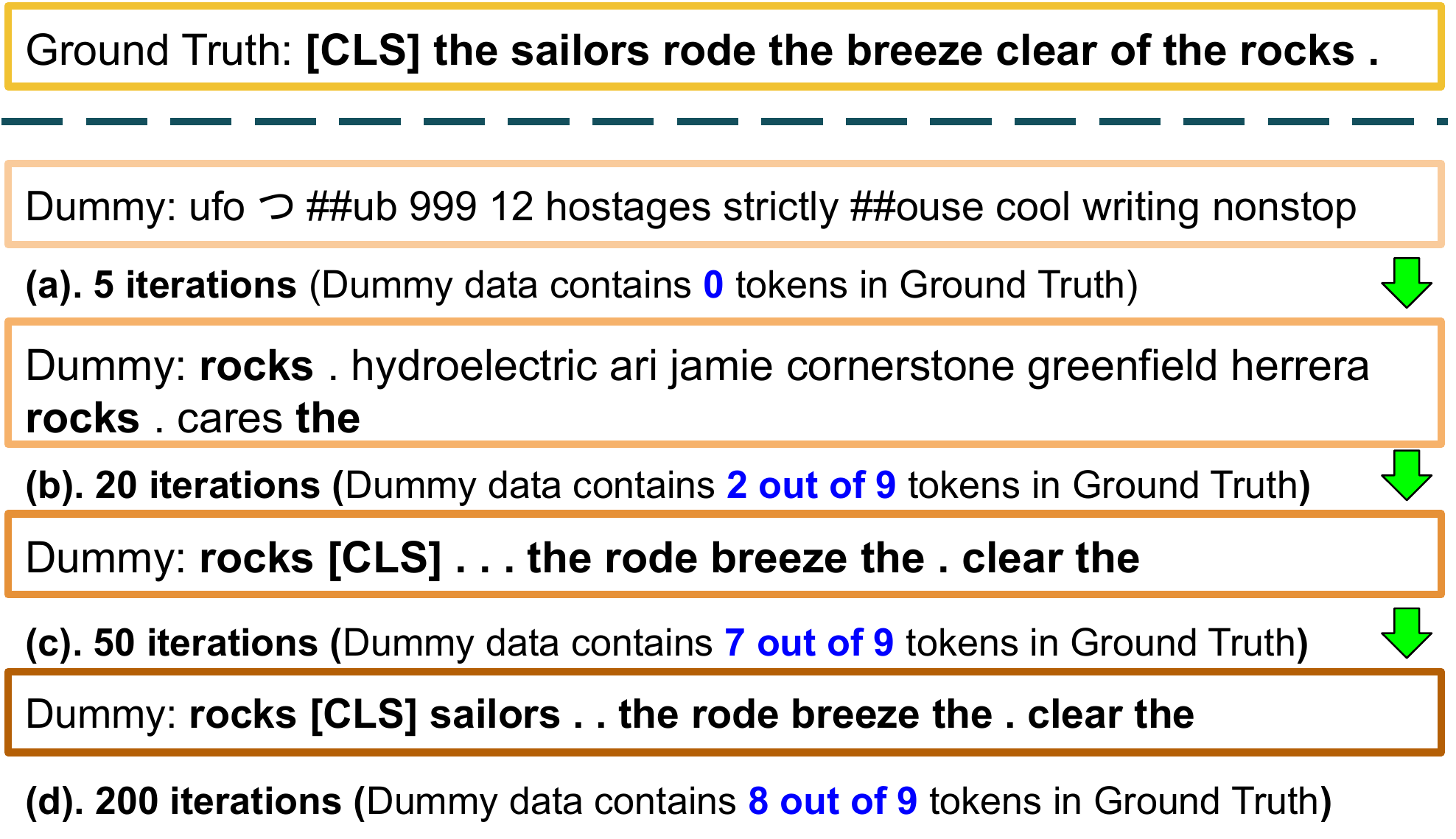}
         \caption{Recover progress of \name{} on a sentence example of CoLA.
         In token level, TAG eventually recovers 8 of 9 tokes (88.89\% Recover Rate) from ground truth which comes from a sentence of CoLA dataset.}
         \label{fig:sentence_level}
\end{figure}

\section{Result Analysis and Visualization}
In this section, we conduct carefully designed experiments to evaluate the proposed \name{} on various datasets mentioned in Section \ref{sec:dataset} using the metrics defined in Section \ref{sec:metric}. 
We have four highlighted results for our evaluation. 

\noindent\textbf{Our algorithm shows stable and distinct convergence for NLP models.} 
Here, we measure the distance between the dummy gradient, $\nabla{{\mathbf{W}}^\prime}$, and the ground truth gradient, $\nabla{{\mathbf{W}}}$, via the aforementioned distance function. We define this distance as the loss of the algorithm. 
We normalize the loss of all selected data at log scale between 0.8 to 1.2 as shown in Fig.~\ref{fig:loss}. The loss is continuously decreasing for different models and we can observe a stable and distinct convergence from the loss curve, especially for the first 200 iterations.

The \name{} attacking process can be visualized in token embeddings level (Fig.~\ref{fig:pca}) and in sentence level (Fig.~\ref{fig:sentence_level}). In tokens embeddings level (Fig.~\ref{fig:pca}), we first reduce the dimension of token embeddings for both dummy input and ground truth by Principal Component Analysis (PCA).
We use the cosine similarity~\cite{li2020sentence} to evaluate similarity of the dimension reduced token embeddings. In Fig.~\ref{fig:pca} (a), the cosine similarity of the token embeddings between dummy data and ground truth is 0.42 at the 5-th iteration which means we can observe a 0.93 cosine similarity of those two token embeddings after 200 iterations. As the number of iterations increases, the increasing cosine similarity indicates that TAG iteratively recovers the data at token embeddings level. 

In sentence level (Fig.~\ref{fig:sentence_level}), we convert the dummy input (\textit{dummy token embeddings}), ${\mathbf{X}^\prime}$, to dummy tokens by the embedding matrix and then a tokenizer can help us to map the tokens with words. In the Fig.~\ref{fig:sentence_level}(a), the dummy data seems random compared to the ground truth (Recover Rate 0\%). After 20 iterations, in the Fig.~\ref{fig:sentence_level}(b), the dummy data contains two tokens (Recover Rate 22.22\%) from the ground truth, \textit{"rocks"} and \textit{"the"}. After 50 iterations, the algorithm has recovered 7 of 9 tokens (Recover Rate 77.78\%) in the ground truth, and one more token has been recovered when it reached 200 iterations (Recover Rate 88.89\%).

\begin{table*}[]
\centering
\scalebox{0.9}{
\begin{tabular}{l|lllll}
\hline
Models & Recover Rate(\%) & ROUGE-1(\%) & ROUGE-2(\%) & ROUGE-L(\%) & Runtime (Seconds) \\ \hline
\textbf{TinyBERT$_{4}$} & 29.45 & 27.07 & 3.12 & 22.41 & 503.24 \\
\textbf{TinyBERT$_{6}$} & 38.37 & 34.95 & 6.54 & 30.87 & 526.01 \\
\textbf{BERT$_{BASE}$} & 40.84 & 41.95 & 7.77 & 38.08 & 1278.62 \\
\textbf{BERT$_{LARGE}$} & 49.62 & 48.67 & 15.03 & 53.09 & 1672.52 \\ \hline 
\end{tabular}}
\caption{The average values of Recover Rate, ROUGE-1, ROUGE-2, ROUGE-L and Runtime. The results are obtained from TinyBERT$_{4}$, TinyBERT$_{6}$, BERT$_{BASE}$ and BERT$_{LARGE}$ on CoLA, SST-2, RTE datasets.}

\label{tab:result comparison}
\end{table*}

\begin{table*}[]
\centering
\scalebox{0.85}{
\label{if_rba312ex}
\begin{tabular}{|l|l|l|r|r|r|r|r|r|r|}
\hline
\multicolumn{1}{|l|}{} & \multicolumn{1}{l|}{Example 1} & \multicolumn{1}{l|}{Example 2}\\ \hline

TAG                                 
& \begin{tabular}[c]{@{}l@{}}
{\bf We} {\bf monitoring the} the {\bf global} 
\\{\bf pandemic} and will and {\bf update} 
\\the {\bf conference plans} of of
\\the {\bf the conference dates} dates.
\end{tabular}

& \begin{tabular}[c]{@{}l@{}}
{\bf The area chairs} {\bf reviewers} reviewers \\will and {\bf area} of conference
\\{\bf broad expertise} expertise {\bf cover} machines 
\\or {\bf cases}
\end{tabular}

\\ \hline
DLG~\cite{zhu2019deep}                               
& \begin{tabular}[c]{@{}l@{}}
{\bf We} we students {\bf monitoring} monitoring {\bf the}
\\{\bf pandemic} and of pandemic plans {\bf plans} \\as needed closer to the {\bf conference dates.}
\end{tabular}                              
& \begin{tabular}[c]{@{}l@{}} 
The we {\bf chairs} chairs written 
\\work {\bf will} will people {\bf expertise} expertise 
\\longer cases {\bf cases}.
\end{tabular}
\\ \hline
Ground Truth                                
& \begin{tabular}[c]{@{}l@{}}
We are monitoring the ongoing global 
\\pandemic and will update the conference plans \\as needed closer to the conference dates.
\end{tabular}                              
& \begin{tabular}[c]{@{}l@{}}
The area chairs and reviewers in each 
\\area will have broad expertise to 
\\cover these cases.
\end{tabular}
\\ \hline
\end{tabular}}
\caption{Recover comparison of DLG~\cite{zhu2019deep} and TAG on sample texts with basic transformer language model. The sentences are selected randomly from online source. Compared to DLG~\cite{zhu2019deep}, \name{} recovers up to 2$\times$ words. }
\label{tab:nlp}
\end{table*}



\noindent\textbf{Larger model leaks more information.} 
Table~\ref{tab:result comparison} summarizes the averaged metrics of TinyBERT$_{4}$, TinyBERT$_{6}$, BERT$_{BASE}$ and BERT$_{LARGE}$ on the mixture of datasets mentioned in Section \ref{sec:dataset}, i.e., RTE, SST-2, and CoLA, with the same vocabulary dictionary. 
According to Table~\ref{tab:model size}, the size of model structure is sequentially increasing from TinyBERT$_{4}$, TinyBERT$_{6}$, BERT$_{BASE}$ to BERT$_{LARGE}$. From Table~\ref{tab:result comparison}, we observe that larger models leak more information than the smaller ones. For Recover Rate, the BERT$_{LARGE}$ leaks 30\% more comparing to the TinyBERT$_{4}$, 20\% more comparing to the TinyBERT$_{6}$ and 10\% more comparing to the BERT$_{BASE}$. A similar result can be found in ROUGE-1. As for ROUGE-2, the information leaked from BERT$_{LARGE}$ is 5$\times$, 2.5$\times$, and 2$\times$ compared to TinyBERT$_{4}$, TinyBERT$_{6}$, and BERT$_{BASE}$,  respectively. For ROUGE-L, the largest model BERT$_{LARGE}$ leaks the most information, which is 2.5$\times$, 1.8$\times$, and 1.5$\times$ larger than TinyBERT$_{4}$, TinyBERT$_{6}$, and BERT$_{BASE}$.

Researchers indicate that \textit{to obtain a better result in NLP, we should use a larger model on a larger dataset} in their paper~\cite{raffel2019exploring}. 
Based on our experiments, smaller NLP models tend to be more resilient against gradient-based attacks. Information and data security could be another dimension adding to the current tradeoffs among accuracy, latency, and memory footprint. 

\noindent\textbf{Larger model requires more time for recovering.} We evaluate the runtime performance of our proposed algorithm among different models under 1,000 iterations. 
A larger model generates more gradients, and in order to recover the data, we need to build the same structure model as the adversarial model to apply our algorithm. 
Hence, in Table~\ref{tab:result comparison}, we can see that runtime increases as the model gets larger. 
BERT$_{LARGE}$ costs 3$\times$ runtime as compared to the TinyBERT$_{4}$, and BERT$_{BASE}$ takes 2.5$\times$ more runtime as compared with TinyBERT$_{4}$.
\begin{figure}[]
\centering
	\includegraphics[width=1\columnwidth]{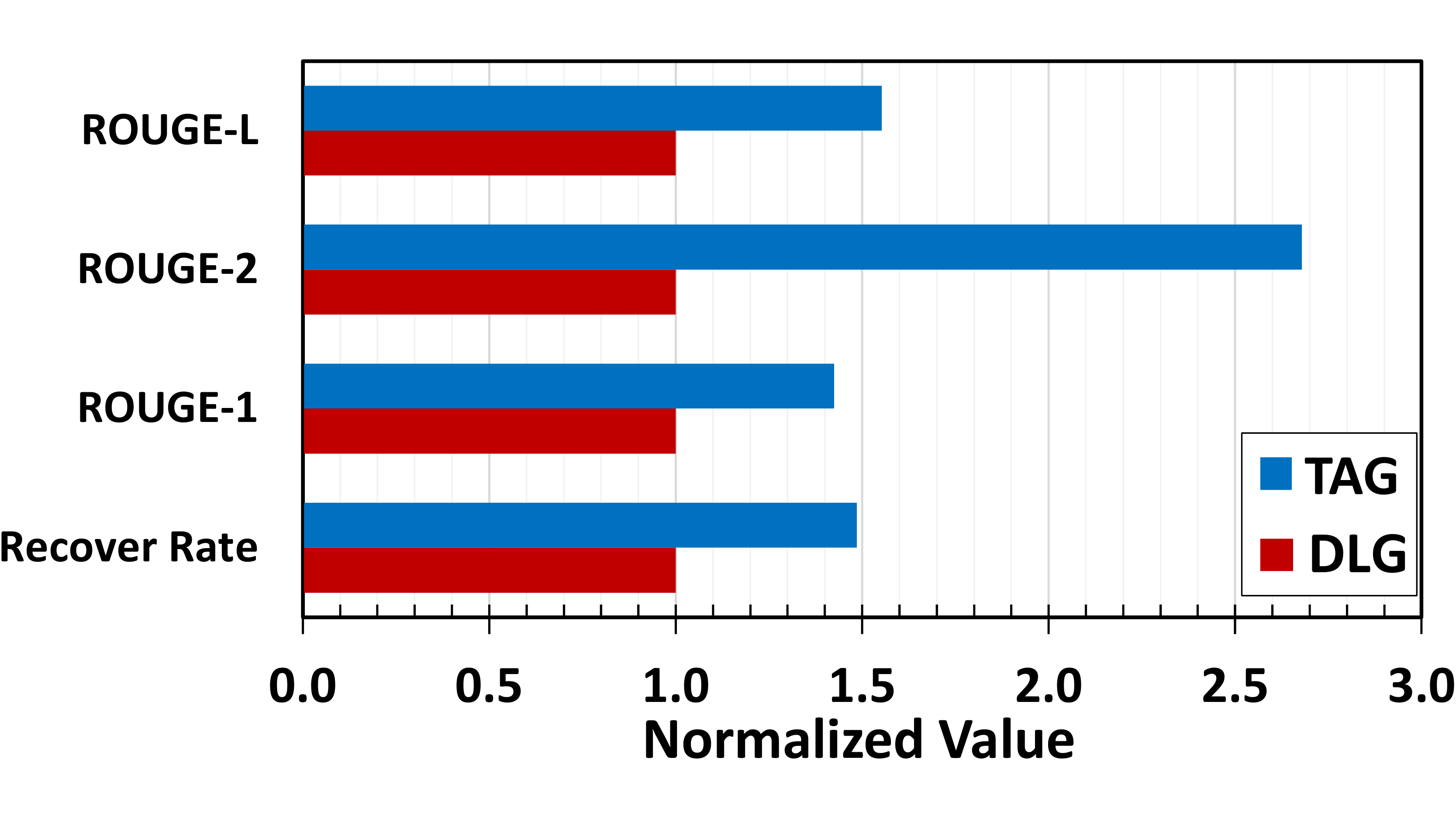}
    \caption{Normalized values of ROUGE-1, ROUGE-2, ROUGE-L, and Recover Rate of \name{} and DLG~\cite{zhu2019deep} (values normalized to DLG metrics). Especially for ROUGE-2, \name{} is more than 2.5$\times$ to DLG~\cite{zhu2019deep}.}
    \label{fig:dlg}
\end{figure}

\noindent\textbf{Our algorithm achieves 2.7$\times$ in ROUGE-2 to prior art.}
We also compare our algorithm with the prior art DLG~\cite{zhu2019deep}. In Table~\ref{tab:nlp}, we apply our algorithm and DLG~\cite{zhu2019deep} on Transformer~\cite{vaswani2017attention} and attack a sentence from online source. Compared to the DLG~\cite{zhu2019deep}, our proposed algorithm recovers more than 2$\times$ words and compares to the ground truth. More importantly, we almost recover all keywords. We further apply \name{} and DLG~\cite{zhu2019deep} on BERT, and evaluate the results on the randomly chosen 100 sentences from CoLA and RTE dataset and calculate the averaged value for each experiment. Fig.~\ref{fig:dlg} shows the results. Compared to DLG~\cite{zhu2019deep}, \name{} demonstrates distinct advantages. For ROUGE-2, the result of \name{} is about 2.7$\times$ to DLG~\cite{zhu2019deep}. As for ROUGE-1, ROUGE-L and Recover Rate, \name{} also takes a 1.5$\times$ advantages to DLG~\cite{zhu2019deep}, which is significant. 

\section{Ablation Studies}
In this section, we conduct ablation experiments over several parameters when we evaluate the results of our algorithm. We change the section of the following factors: the weight distributions, the pre-trained weight, the length of the sentence data, and the size of the vocabulary dictionary.

\subsection{Effects of weight distributions}
We evaluate the effects of weight distributions by different distributions and different standard deviations of the distributions. As shown in Table~\ref{tab:weight_distribution}, we use the TinyBERT$_{6}$ model and choose sample data from CoLA to apply different weight distributions. For normal distribution with mean as 0, TAG can recover half words from the sentence when standard deviation is 0.01 while it can only recover one of three words from the sentence with a 0.03 standard deviation. For the uniform distribution weight initialization, the results show that TAG is able to recover more with larger initialization range.
\begin{table}[t]
\centering
\scalebox{0.65}{
\begin{tabular}{lllllll}
\hline
\multirow{2}{*}{Weight Distribution} & \multicolumn{3}{l}{\begin{tabular}[c]{@{}l@{}}Uniform \\ (Initializer Range)\end{tabular}} & \multicolumn{3}{l}{\begin{tabular}[c]{@{}l@{}}Normal (Mean=0) \\ (Standard Deviation)\end{tabular}} \\ \cline{2-7} 
 & \textbf{$\pm$0.01} & \textbf{$\pm$0.02} & \textbf{$\pm$0.03} & \textbf{0.01} & \textbf{0.02} & \textbf{0.03} \\ \hline
Recover Rate(\%) & 36.21 & 52.17 & 60.25 & 50.12 & 41.57 & 33.33 \\
ROUGE-1(\%) & 39.39 & 44.27 & 60.98 & 54.54 & 45.56 & 35.71 \\
ROUGE-2(\%) & 14.54 & 15.09 & 23.63 & 30.00 & 1.01 & 0 \\
ROUGE-L(\%) & 44.39 & 46.98 & 57.43 & 66.66 & 40.01 & 37.01 \\ \hline
\end{tabular}}
\caption{Recover Rate, ROUGE-1, ROUGE-2, ROUGE-L values of \name{} with TinyBERT$_{6}$ uniform distribution and normal distribution on sample sentence from CoLA under.}
\label{tab:weight_distribution}
\end{table}
\begin{table}[t]
\centering
\scalebox{0.8}{
\begin{tabular}{l|ll|ll}
\hline
 Models & \multicolumn{2}{l|}{Pre-trained Model} & \multicolumn{2}{l}{Initialized Model} \\ \hline
Datasets & \textbf{CoLA} & \textbf{SST-2} & \textbf{CoLA} & \textbf{SST-2} \\ \hline
Recover Rate(\%) & 48.76 & 43.85 & 34.13 & 33.82 \\
ROUGE-1(\%) & 45.68 & 36.40 & 30.84 & 30.74 \\
ROUGE-2(\%) & 8.01 & 4.26 & 6.41 & 5.45 \\
ROUGE-L(\%) & 37.61 & 32.95 & 26.80 & 26.42 \\ \hline
\end{tabular}}
\caption{Recover Rate, ROUGE-1, ROUGE-2, ROUGE-L values of \name{} with TinyBERT${_6}$ on weight from pre-trained model and normal initialized model.}
\label{tab:pretrain_initialized}
\end{table}
\subsection{Effects of weights from pre-trained model}
We evaluate our proposed algorithm on the effects of weights from pre-trained model on two different datasets, CoLA and SST-2. In this experiment, we choose the TinyBERT$_{6}$ model and download the pre-trained version from GitHub and also initialize this model using normal distribution with mean as 0 and standard deviation as 0.02. In Table~\ref{tab:pretrain_initialized}, for the CoLA dataset, pre-trained model demonstrates 1.5$\times$ better than the initialized model. Overall, the pre-trained model shows a better result than the initialized model. We consider that the pre-trained model may contain more information during the training process than the initialized model.

\begin{table}[t]
\centering
\scalebox{0.75}{
\begin{tabular}{l|llll}
\hline
Datasets & \begin{tabular}[c]{@{}l@{}}Recover \\ Rate(\%)\end{tabular} & R-1(\%) & R-2(\%) & R-L(\%) \\ \hline
\textbf{RTE($\sim$50 words)} & 22.70 & 13.40 & 1.09 & 11.29 \\
\textbf{CoLA($\sim$10 words)} & 34.13 & 30.84 & 6.41 & 26.80 \\ \hline
\end{tabular}}
\caption{Recover Rate, ROUGE-1 (R-1), ROUGE-2 (R-2), and ROUGE-L (R-L) values of TAG on comparison of different length sentences from RTE and CoLA.}
\label{tab:differnt_dataset}
\end{table}

\begin{table}[h!]
\resizebox{0.99\columnwidth}{!}{
\begin{tabular}{l|cc|c}
\hline
Vocabulary & Small-Scale & Medium-Scale & Ratio\\ \hline
Total \# of Tokens & 21,128 & 30,522 & 0.69 \\
Recover Rate(\%) & 54.61 &  34.13 & 1.60 \\
ROUGE-1(\%) & 54.87 & 30.84 & 1.78 \\
ROUGE-2(\%) & 11.83 & 6.41 & 1.85 \\
ROUGE-L(\%) & 47.40 & 26.80 & 1.77 \\
\hline
\end{tabular}
}
\caption{Recover Rate, ROUGE-1, ROUGE-2, and ROUGE-L values of TAG on comparison of different scales of vocabulary dictionaries.}
\label{tab:vocab}
\end{table}


\subsection{Performance on different datasets}
To evaluate the effects of different sentence length to our proposed algorithm, we conduct experiments on datasets: RTE and CoLA. RTE is a dataset that contains longer sentences than CoLA. We choose sentences to contain more than 50 words from RTE, while sentences within ten words from CoLA as the input data for this experiment. We choose the TinyBERT$_{6}$ model with initialized normal distributed weight for this experiment. In Table~\ref{tab:differnt_dataset}, the results from CoLA are better than RTE, especially for ROUGE family. The ROUGE-1 and ROUGE-2 of CoLA are 3$\times$ better than RTE, and ROUGE-L is 2.5$\times$ better than RTE.



\subsection{Effects of vocabulary dictionary}
To evaluate the effects of vocabulary scale, we choose a small scale vocabulary from~\cite{cui2019pre} and a medium scale vocabulary from BERT~\cite{devlin2019bert}. 
The total numbers of tokens in the small and medium vocabularies are 21,128 and 30,522, respectively. We use TinyBERT$_{6}$ model on CoLA and only alter the vocabulary. In Table~\ref{tab:vocab}, we observe that the smaller vocabulary size may result in more leakage while the larger one leaks less. For the smaller vocabulary size, the result is more than 1.6$\times$ improvement compared to the larger one in terms of all evaluation metrics.


\section{Conclusion}
In this work, we propose, \name{}, Transformer Attack from Gradient framework with an adversary algorithm to recover private text data from the transformer model's gradients. We demonstrate that \name{} addresses private information like name is likely to be leaked in transformer-based model. We develop a set of metrics to evaluate the effectiveness of the proposed attack algorithm quantitatively. Our experiments show that \name{} works well on more different weight distributions in recovering private training data on Transformer, TinyBERT$_{4}$, TinyBERT$_{6}$, BERT$_{BASE}$, and BERT$_{LARGE}$ using GLUE benchmark, and achieves 1.5$\times$ Recover Rate and 2.5$\times$ ROUGE-2 over prior methods without the need of ground truth label. \name{} can obtain up to 88.9$\%$ tokens and up to 0.93 cosine similarity in token embeddings from private training data by attacking gradients on CoLA dataset.
We hope the proposed \name{} will shed some light on the privacy leakage problem in Transformer-based NLP models.

\section{Acknowledgements}
This research is supported in part by the National Science Foundation (NSF) Grants 1743418 and 1843025. Hang Liu is in part supported by NSF CRII Award No. 2000722 and CAREER Award No. 204610.
\bibliographystyle{acl_natbib}
\bibliography{custom,anthology}

\begin{thebibliography}{34}
\expandafter\ifx\csname natexlab\endcsname\relax\def\natexlab#1{#1}\fi

\bibitem[{Bao et~al.(2020)Bao, Dong, Wei, Wang, Yang, Liu, Wang, Piao, Gao,
  Zhou et~al.}]{bao2020unilmv2}
Hangbo Bao, Li~Dong, Furu Wei, Wenhui Wang, Nan Yang, Xiaodong Liu, Yu~Wang,
  Songhao Piao, Jianfeng Gao, Ming Zhou, et~al. 2020.
\newblock Unilmv2: Pseudo-masked language models for unified language model
  pre-training.
\newblock \emph{arXiv preprint arXiv:2002.12804}.

\bibitem[{Baruch et~al.(2019)Baruch, Baruch, and
  Goldberg}]{NEURIPS2019_ec1c5914}
Gilad Baruch, Moran Baruch, and Yoav Goldberg. 2019.
\newblock A little is enough: Circumventing defenses for distributed learning.
\newblock In \emph{Advances in Neural Information Processing Systems}. Curran
  Associates, Inc.

\bibitem[{Brown et~al.(2020)Brown, Mann, Ryder, Subbiah, Kaplan, Dhariwal,
  Neelakantan, Shyam, Sastry, Askell et~al.}]{GPT-3}
Tom~B Brown, Benjamin Mann, Nick Ryder, Melanie Subbiah, Jared Kaplan, Prafulla
  Dhariwal, Arvind Neelakantan, Pranav Shyam, Girish Sastry, Amanda Askell,
  et~al. 2020.
\newblock Language models are few-shot learners.
\newblock \emph{arXiv preprint arXiv:2005.14165}.

\bibitem[{Chen et~al.(2020)Chen, Jia, and Qi}]{chen2020improved}
Si~Chen, Ruoxi Jia, and Guo-Jun Qi. 2020.
\newblock \href {http://arxiv.org/abs/2010.04092} {Improved techniques for
  model inversion attacks}.

\bibitem[{Cui et~al.(2019)Cui, Che, Liu, Qin, Yang, Wang, and Hu}]{cui2019pre}
Yiming Cui, Wanxiang Che, Ting Liu, Bing Qin, Ziqing Yang, Shijin Wang, and
  Guoping Hu. 2019.
\newblock Pre-training with whole word masking for chinese bert.
\newblock \emph{arXiv preprint arXiv:1906.08101}.

\bibitem[{Dagan et~al.(2005)Dagan, Glickman, and Magnini}]{RTE-dagan2005pascal}
Ido Dagan, Oren Glickman, and Bernardo Magnini. 2005.
\newblock The pascal recognising textual entailment challenge.
\newblock In \emph{Machine Learning Challenges Workshop}, pages 177--190.
  Springer.

\bibitem[{Das et~al.(2016)Das, Avancha, Mudigere, Vaidynathan, Sridharan,
  Kalamkar, Kaul, and Dubey}]{das2016distributed}
Dipankar Das, Sasikanth Avancha, Dheevatsa Mudigere, Karthikeyan Vaidynathan,
  Srinivas Sridharan, Dhiraj Kalamkar, Bharat Kaul, and Pradeep Dubey. 2016.
\newblock Distributed deep learning using synchronous stochastic gradient
  descent.
\newblock \emph{arXiv preprint arXiv:1602.06709}.

\bibitem[{Dean et~al.(2012)Dean, Corrado, Monga, Chen, Devin, Mao, Ranzato,
  Senior, Tucker, Yang, Le, and Ng}]{NIPS2012_6aca9700}
Jeffrey Dean, Greg Corrado, Rajat Monga, Kai Chen, Matthieu Devin, Mark Mao,
  Marc\textquotesingle~aurelio Ranzato, Andrew Senior, Paul Tucker, Ke~Yang,
  Quoc Le, and Andrew Ng. 2012.
\newblock Large scale distributed deep networks.
\newblock In \emph{Advances in Neural Information Processing Systems},
  volume~25. Curran Associates, Inc.

\bibitem[{Devlin et~al.(2019)Devlin, Chang, Lee, and
  Toutanova}]{devlin2019bert}
Jacob Devlin, Ming-Wei Chang, Kenton Lee, and Kristina Toutanova. 2019.
\newblock \href {http://arxiv.org/abs/1810.04805} {Bert: Pre-training of deep
  bidirectional transformers for language understanding}.

\bibitem[{Fredrikson et~al.(2015)Fredrikson, Jha, and
  Ristenpart}]{fredrikson2015model}
Matt Fredrikson, Somesh Jha, and Thomas Ristenpart. 2015.
\newblock Model inversion attacks that exploit confidence information and basic
  countermeasures.
\newblock In \emph{Proceedings of the 22nd ACM SIGSAC Conference on Computer
  and Communications Security}, pages 1322--1333.

\bibitem[{Geiping et~al.(2020)Geiping, Bauermeister, Dr\"{o}ge, and
  Moeller}]{NEURIPS2020_c4ede56b}
Jonas Geiping, Hartmut Bauermeister, Hannah Dr\"{o}ge, and Michael Moeller.
  2020.
\newblock \href
  {https://proceedings.neurips.cc/paper/2020/file/c4ede56bbd98819ae6112b20ac6bf145-Paper.pdf}
  {Inverting gradients - how easy is it to break privacy in federated
  learning?}
\newblock In \emph{Advances in Neural Information Processing Systems},
  volume~33, pages 16937--16947. Curran Associates, Inc.

\bibitem[{Goodfellow et~al.(2014)Goodfellow, Pouget-Abadie, Mirza, Xu,
  Warde-Farley, Ozair, Courville, and Bengio}]{goodfellow2014generative}
Ian Goodfellow, Jean Pouget-Abadie, Mehdi Mirza, Bing Xu, David Warde-Farley,
  Sherjil Ozair, Aaron Courville, and Yoshua Bengio. 2014.
\newblock Generative adversarial nets.
\newblock In \emph{Advances in neural information processing systems}, pages
  2672--2680.

\bibitem[{Gurevin et~al.(2021)Gurevin, Bragin, Ding, Zhou, Pepin, Li, and
  Miao}]{Gurevin2021Enabling}
Deniz Gurevin, Mikhail Bragin, Caiwen Ding, Shanglin Zhou, Lynn Pepin, Bingbing
  Li, and Fei Miao. 2021.
\newblock \href {https://doi.org/10.24963/ijcai.2021/344} {Enabling
  retrain-free deep neural network pruning using surrogate lagrangian
  relaxation}.
\newblock In \emph{Proceedings of the Thirtieth International Joint Conference
  on Artificial Intelligence, {IJCAI-21}}, pages 2497--2504. International
  Joint Conferences on Artificial Intelligence Organization.
\newblock Main Track.

\bibitem[{Hendrycks and Gimpel(2016)}]{hendrycks2016gaussian}
Dan Hendrycks and Kevin Gimpel. 2016.
\newblock Gaussian error linear units (gelus).
\newblock \emph{arXiv preprint arXiv:1606.08415}.

\bibitem[{Jiao et~al.(2020)Jiao, Yin, Shang, Jiang, Chen, Li, Wang, and
  Liu}]{jiao2020tinybert}
Xiaoqi Jiao, Yichun Yin, Lifeng Shang, Xin Jiang, Xiao Chen, Linlin Li, Fang
  Wang, and Qun Liu. 2020.
\newblock \href {http://arxiv.org/abs/1909.10351} {Tinybert: Distilling bert
  for natural language understanding}.

\bibitem[{Li et~al.(2020)Li, Zhou, He, Wang, Yang, and Li}]{li2020sentence}
Bohan Li, Hao Zhou, Junxian He, Mingxuan Wang, Yiming Yang, and Lei Li. 2020.
\newblock On the sentence embeddings from pre-trained language models.
\newblock \emph{arXiv preprint arXiv:2011.05864}.

\bibitem[{Li et~al.(2014{\natexlab{a}})Li, Andersen, Park, Smola, Ahmed,
  Josifovski, Long, Shekita, and Su}]{10.5555/2685048.2685095}
Mu~Li, David~G. Andersen, Jun~Woo Park, Alexander~J. Smola, Amr Ahmed, Vanja
  Josifovski, James Long, Eugene~J. Shekita, and Bor-Yiing Su.
  2014{\natexlab{a}}.
\newblock Scaling distributed machine learning with the parameter server.
\newblock In \emph{Proceedings of the 11th USENIX Conference on Operating
  Systems Design and Implementation}, OSDI'14, page 583–598, USA. USENIX
  Association.

\bibitem[{Li et~al.(2014{\natexlab{b}})Li, Andersen, Smola, and
  Yu}]{NIPS2014_1ff1de77}
Mu~Li, David~G Andersen, Alexander~J Smola, and Kai Yu. 2014{\natexlab{b}}.
\newblock Communication efficient distributed machine learning with the
  parameter server.
\newblock In \emph{Advances in Neural Information Processing Systems},
  volume~27. Curran Associates, Inc.

\bibitem[{Lin(2004)}]{lin-2004-rouge}
Chin-Yew Lin. 2004.
\newblock \href {https://www.aclweb.org/anthology/W04-1013} {{ROUGE}: A package
  for automatic evaluation of summaries}.
\newblock In \emph{Text Summarization Branches Out}, pages 74--81, Barcelona,
  Spain. Association for Computational Linguistics.

\bibitem[{Lin et~al.(2020)Lin, Wang, Li, Deng, Wang, and Ding}]{Lin2020ESMFLEA}
Sheng-Jie Lin, Chenghong Wang, Hongjia Li, Jieren Deng, Yanzhi Wang, and Caiwen
  Ding. 2020.
\newblock Esmfl: Efficient and secure models for federated learning.
\newblock \emph{ArXiv}, abs/2009.01867.

\bibitem[{Liu et~al.(2019)Liu, Ott, Goyal, Du, Joshi, Chen, Levy, Lewis,
  Zettlemoyer, and Stoyanov}]{liu2019roberta}
Yinhan Liu, Myle Ott, Naman Goyal, Jingfei Du, Mandar Joshi, Danqi Chen, Omer
  Levy, Mike Lewis, Luke Zettlemoyer, and Veselin Stoyanov. 2019.
\newblock Roberta: A robustly optimized bert pretraining approach.
\newblock \emph{arXiv preprint arXiv:1907.11692}.

\bibitem[{Melis et~al.(2019)Melis, Song, De~Cristofaro, and
  Shmatikov}]{melis2019exploiting}
Luca Melis, Congzheng Song, Emiliano De~Cristofaro, and Vitaly Shmatikov. 2019.
\newblock Exploiting unintended feature leakage in collaborative learning.
\newblock In \emph{2019 IEEE Symposium on Security and Privacy (SP)}, pages
  691--706. IEEE.

\bibitem[{Peters et~al.(2018)Peters, Neumann, Iyyer, Gardner, Clark, Lee, and
  Zettlemoyer}]{peters2018deep}
Matthew~E. Peters, Mark Neumann, Mohit Iyyer, Matt Gardner, Christopher Clark,
  Kenton Lee, and Luke Zettlemoyer. 2018.
\newblock \href {http://arxiv.org/abs/1802.05365} {Deep contextualized word
  representations}.

\bibitem[{Raffel et~al.(2019{\natexlab{a}})Raffel, Shazeer, Roberts, Lee,
  Narang, Matena, Zhou, Li, and Liu}]{2019t5}
Colin Raffel, Noam Shazeer, Adam Roberts, Katherine Lee, Sharan Narang, Michael
  Matena, Yanqi Zhou, Wei Li, and Peter~J Liu. 2019{\natexlab{a}}.
\newblock Exploring the limits of transfer learning with a unified text-to-text
  transformer.
\newblock \emph{arXiv}, pages arXiv--1910.

\bibitem[{Raffel et~al.(2019{\natexlab{b}})Raffel, Shazeer, Roberts, Lee,
  Narang, Matena, Zhou, Li, and Liu}]{raffel2019exploring}
Colin Raffel, Noam Shazeer, Adam Roberts, Katherine Lee, Sharan Narang, Michael
  Matena, Yanqi Zhou, Wei Li, and Peter~J Liu. 2019{\natexlab{b}}.
\newblock Exploring the limits of transfer learning with a unified text-to-text
  transformer.
\newblock \emph{arXiv preprint arXiv:1910.10683}.

\bibitem[{Sanh et~al.(2019)Sanh, Debut, Chaumond, and
  Wolf}]{sanh2019distilbert}
Victor Sanh, Lysandre Debut, Julien Chaumond, and Thomas Wolf. 2019.
\newblock Distilbert, a distilled version of bert: smaller, faster, cheaper and
  lighter.
\newblock \emph{arXiv preprint arXiv:1910.01108}.

\bibitem[{Socher et~al.(2013)Socher, Perelygin, Wu, Chuang, Manning, Ng, and
  Potts}]{socher-etal-2013-recursive-SST-2}
Richard Socher, Alex Perelygin, Jean Wu, Jason Chuang, Christopher~D Manning,
  Andrew~Y Ng, and Christopher Potts. 2013.
\newblock Recursive deep models for semantic compositionality over a sentiment
  treebank.
\newblock In \emph{Proceedings of the 2013 conference on empirical methods in
  natural language processing}, pages 1631--1642.

\bibitem[{Vaswani et~al.(2017)Vaswani, Shazeer, Parmar, Uszkoreit, Jones,
  Gomez, Kaiser, and Polosukhin}]{vaswani2017attention}
Ashish Vaswani, Noam Shazeer, Niki Parmar, Jakob Uszkoreit, Llion Jones,
  Aidan~N Gomez, {\L}ukasz Kaiser, and Illia Polosukhin. 2017.
\newblock Attention is all you need.
\newblock In \emph{Advances in neural information processing systems}, pages
  5998--6008.

\bibitem[{Wang et~al.(2019)Wang, Singh, Michael, Hill, Levy, and
  Bowman}]{wang2019glue}
Alex Wang, Amanpreet Singh, Julian Michael, Felix Hill, Omer Levy, and
  Samuel~R. Bowman. 2019.
\newblock \href {http://arxiv.org/abs/1804.07461} {Glue: A multi-task benchmark
  and analysis platform for natural language understanding}.

\bibitem[{Wang et~al.(2021)Wang, Wang, Wang, Zhou, Liu, Bi, Ding, and
  Rajasekaran}]{ijcai2021-432}
Yijue Wang, Chenghong Wang, Zigeng Wang, Shanglin Zhou, Hang Liu, Jinbo Bi,
  Caiwen Ding, and Sanguthevar Rajasekaran. 2021.
\newblock \href {https://doi.org/10.24963/ijcai.2021/432} {Against membership
  inference attack: Pruning is all you need}.
\newblock In \emph{Proceedings of the Thirtieth International Joint Conference
  on Artificial Intelligence, {IJCAI-21}}, pages 3141--3147. International
  Joint Conferences on Artificial Intelligence Organization.
\newblock Main Track.

\bibitem[{Warstadt et~al.(2019)Warstadt, Singh, and Bowman}]{cola2019}
Alex Warstadt, Amanpreet Singh, and Samuel~R. Bowman. 2019.
\newblock \href {http://arxiv.org/abs/1805.12471} {Neural network acceptability
  judgments}.

\bibitem[{Wu et~al.(2020)Wu, Zheng, Dou, Chen, Deng, Chen, Xu, Gao, Li, Wang,
  Xiao, Xie, Wang, and Xu}]{10.1093/bib/bbaa090}
Xin Wu, Hao Zheng, Zuochao Dou, Feng Chen, Jieren Deng, Xiang Chen, Shengqian
  Xu, Guanmin Gao, Mengmeng Li, Zhen Wang, Yuhui Xiao, Kang Xie, Shuang Wang,
  and Huji Xu. 2020.
\newblock \href {https://doi.org/10.1093/bib/bbaa090} {{A novel
  privacy-preserving federated genome-wide association study framework and its
  application in identifying potential risk variants in ankylosing
  spondylitis}}.
\newblock \emph{Briefings in Bioinformatics}, 22(3).
\newblock Bbaa090.

\bibitem[{Yang et~al.(2019)Yang, Dai, Yang, Carbonell, Salakhutdinov, and
  Le}]{yang2019xlnet}
Zhilin Yang, Zihang Dai, Yiming Yang, Jaime Carbonell, Russ~R Salakhutdinov,
  and Quoc~V Le. 2019.
\newblock Xlnet: Generalized autoregressive pretraining for language
  understanding.
\newblock In \emph{Advances in neural information processing systems}, pages
  5754--5764.

\bibitem[{Zhu et~al.(2019)Zhu, Liu, and Han}]{zhu2019deep}
Ligeng Zhu, Zhijian Liu, and Song Han. 2019.
\newblock Deep leakage from gradients.
\newblock In \emph{Advances in Neural Information Processing Systems}, pages
  14774--14784.

\end{thebibliography}
\end{document}